 \documentclass[prb,floats,showpacs,onecolum,twocolumn]{revtex4-1}

\usepackage[english]{babel}
\usepackage{amsmath}
\usepackage{amssymb, comment}
\usepackage[]{graphicx}
\usepackage{times}
\usepackage{bm}
\usepackage{units}
\usepackage{braket}
\usepackage[version=3]{mhchem}
\newcommand{\eps}{\varepsilon}
\usepackage{bbm}

\newcommand{\up}{\ensuremath{\uparrow}}
\newcommand{\down}{\ensuremath{\downarrow}}

\newcommand{\I}{\ensuremath{\mathbbm{1}}}

\usepackage{color}

\renewcommand{\eqref}[1]{Eq. (\ref{#1})}

\begin{document}
\title{Brightening of spin- and momentum-dark excitons in transition metal dichalcogenides }

\author{Maja Feierabend, Samuel Brem, August Ekman, and Ermin Malic}
\address{Chalmers University of Technology, Department of 
Physics, 412 96 Gothenburg, Sweden}


\begin{abstract}
Monolayer transition metal dichalcogenides (TMDs) have been in focus of current research, among others due to their remarkable exciton landscape consisting of bright and dark excitonic states. Although dark excitons are not directly visible in optical spectra, they have a large impact on  exciton dynamics  and hence their understanding is crucial for potential TMD-based applications. Here, we study brightening mechanisms of dark excitons via interaction with phonons and in-plane magnetic fields.  We show clear signatures of momentum- and spin-dark excitons  in WS$_2$, WSe$_2$ and MoS$_2$, while the photoluminescence of MoSe$_2$ is only determined by the bright exciton. In particular, we reveal the mechanism behind the brightening of states that are both spin- \textit{and} momentum-dark in MoS$_2$.  Our results are in good agreement with recent experiments and contribute to a better microscopic understanding of the exciton landscape in TMDs. 
\end{abstract}

\maketitle
\begin{figure}[t]
\includegraphics[width=\linewidth]{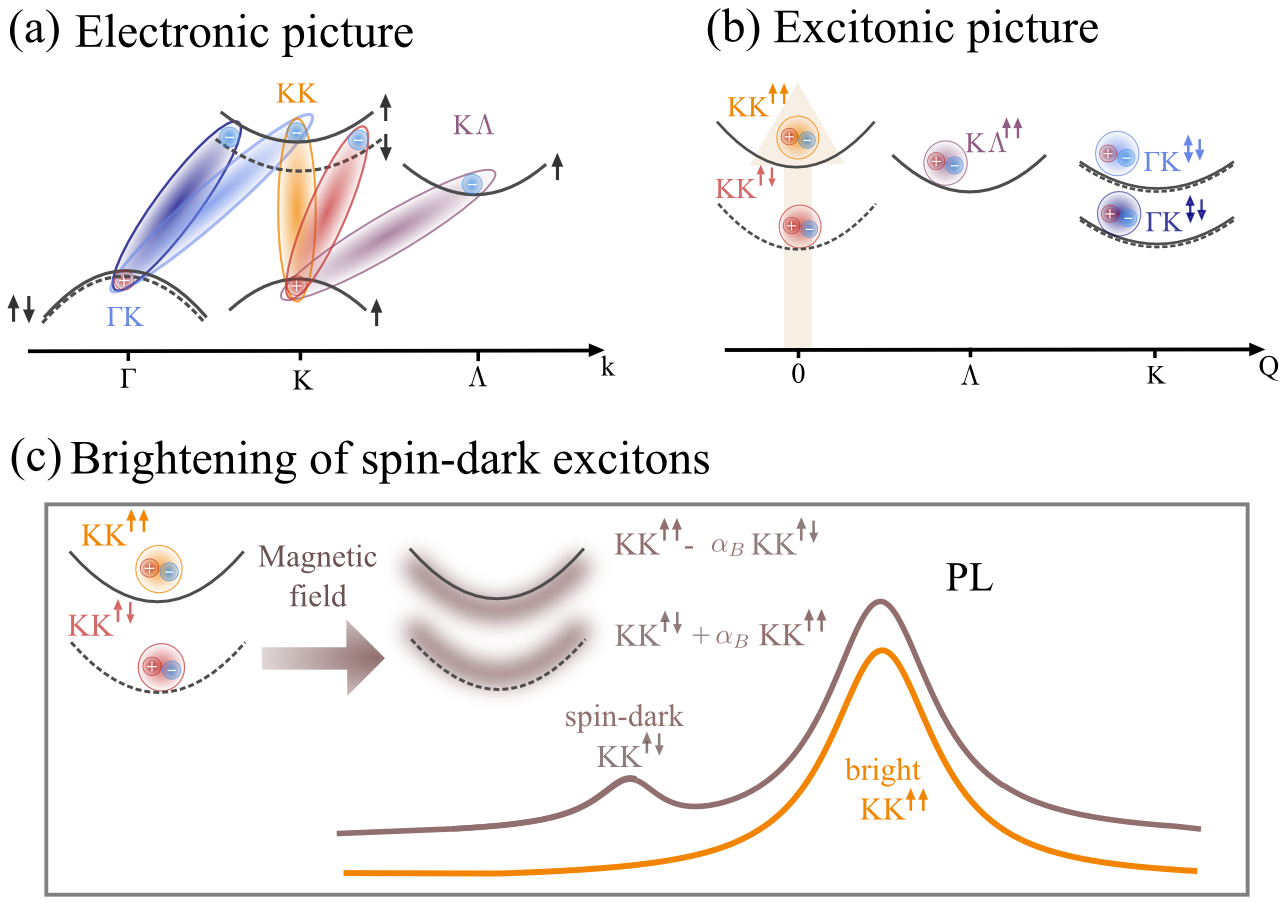} 
    \caption{\textbf{Exciton dispersion and influence of a magnetic field.} (a) Electronic band structure around the high-symmetry $\Gamma$, K, $\Lambda$ points including spin-orbit coupling (solid/dashed lines represent spin-up/down configurations).  (b) Corresponding exciton picture with bright KK$^{\up\up}$,  spin-dark KK$^{\up\down}$ and momentum-dark $\text{K}\Lambda^{\up\up}$, $\Gamma \text{K}^{\up\up}$ as well as spin- \textit{and} momentum-dark $\Gamma \text{K}^{\up\down}$ excitons. Note that this is a schematic figure and the exact position of the corresponding valleys depend on the TMD material. (c) Brightening of spin-dark excitons due to magnetic field-induced spin-mixing of KK$^{\up\up}$ and KK$^{\up\down}$ states. The factor $\alpha_B=\frac{g\mu_BB}{2\Delta}$ represents a mixing parameter that determines whether an additional peak appears in PL spectra stemming from the spin-dark exciton.}
   \label{figure1}
\end{figure}

%
%
Transition metal dichalcogenides (TMDs) exhibit a number of fundamentally interesting and technologically promising properties \cite{alexeyReview,erminnpj,merkl2019ultrafast}.
Their electronic band structure consists of multiple minima and maxima in the  valence and conduction band, which - combined with the strong Coulomb interaction - leads to a variety of  exciton states, cf. Fig. \ref{figure1}(a) \cite{THeinz,gunnar_prb,alexeyReview}. Intervalley excitons consisting of electrons and holes located in different valleys (K, $\Lambda, \Gamma$), are momentum-dark since photons cannot provide the required momentum  necessary for an indirect recombination \cite{selig2016excitonic}.  Furthermore, the spin-orbit coupling gives rise to pronounced spin-splitting in both conduction and valence bands \cite{andor,yu2019exploration} leading to spin-allowed (same spin for valence and conduction band) and spin-dark (different spin) states. 
 While bright excitons, consisting of Coulomb-bound electrons and holes in the same valley with the same spin, can be directly activated by light and have been extensively investigated in literature \cite{THeinz,Chernikov2014, gunnar_prb,arora2015excitonic,erminnpj}, spin- and/or momentum-dark excitons need an additional brightening mechanism to be visible in optical spectra \cite{maja_sensor,molas2017brightening}. Recently, signatures of spin-dark excitons have been observed in experiments with large aperture even in the absence of a magnetic field \cite{li2019emerging} and hence spin-dark excitons are rather \textit{darkish}, since they exhibit an out of plane dipole.
 
 Dark excitons are highly interesting for TMD research, as they can lie energetically below  bright excitons \cite{selig2018dark,malic2018dark,deilmann2019finite}(Fig. \ref{figure1}(b)) and hence have a significant impact on non-equilibrium dynamics as well as optical response of these materials.   
Different mechanisms can principally brighten up  dark exciton states. This includes in-plane magnetic fields, which mix the spin states making spin-dark excitons visible \cite{molas2017brightening,zhang2017magnetic}. Phonons, disorder or  molecules provide an additional center-of-mass momentum to activate momentum-dark excitons \cite{zhang2015experimental,lindlau2018role,lindlau2017identifying,zhou2017probing,feierabend2018molecule,brem2019phonon}. 
In this work, we present a microscopic approach allowing us to investigate the possibility to brighten up states that are both spin- and momentum-dark. 
Our work is motivated by an experimental study observing a yet unidentified low-energy peak in MoS$_2$ monolayers in presence of an in-plane magnetic field. 
While  brightening of spin-dark excitons in tungsten-based TMDs has been  well understood \cite{robert2017fine,wang2017plane,vasconcelos2018dark,peng2019distinctive,baranowski2017dark} ,
only little is known about spin- \textit{and} momentum-dark excitons in MoS$_2$.
Including a magnetic field in our equation-of-motion approach, we find a field-induced mixing of spin-up and spin-down states, which activates the originally spin-dark exciton resulting in an additional peak in optical spectra, cf. Fig \ref{figure1}(c). Including phonon-assisted optical transitions on the same microscopic footing, we investigate the possibility to brighten up even states that are both spin- \textit{and}  momentum-dark. \\

%
%
{\noindent\textbf{Theoretical approach:}} To obtain a microscopic access to the optical response of TMDs after an optical excitation, we  apply the density matrix formalism with semiconductor Bloch equations in its core \cite{Kochbuch, Kira2006,carbonbuch,kadi14}. The particular goal of this work is to describe many-particle mechanisms brightening up momentum- and spin-dark states. The intensity of photoluminescence (PL) can be expressed as \cite{thranhardt2000quantum,feierabend2018molecule,brem2019phonon}
\begin{equation}
  \label{eq:pl_intensity}
  I(\omega_q)
  \propto \hbar \omega_q \partial_t \braket{c_q^\dagger c_q}
  \propto 
  \Im\left[
    \braket{c_q^\dagger X_q^b}  \right].
\end{equation}
The PL is given by the time derivative of the photon density  $\braket{c_q^\dagger c_q}$ that is determined by  the photon-assisted polarization $\braket{c_q^\dagger X_q^b}$  corresponding to the recombination of an exciton ($X_q^b$) under emission
of a photon ($c_q^\dagger$). Here, we have introduced
 the photon creation and annihilation operators $ c_q^{\dagger}, c_q$ and the exciton annihilation operator $X_q^b$ that will be defined below. 
The introduced PL equation only describes direct radiative recombination processes and thus contains only signatures from the bright exciton. To include also possible features stemming from dark excitons via higher order processes,  we have to extend the PL equation by implementing phonon-assisted radiative recombination processes and the impact of a magnetic field. 

To account for excitonic effects, which are dominant in TMD monolayers \cite{Chernikov2014, gunnar_prb,arora2015excitonic,erminnpj,raja2019dielectric}, it is convenient to project the many-particle system into an excitonic basis. Following  Katsch et al. \cite{katsch2018theory}, we introduce the exciton operator
$
X_Q = \sum_{q} \varphi^{*}_{q} a^{c\dagger}_{q-\alpha Q} a^v_{q+\beta Q},
$
which includes the exciton wavefunction $\varphi^*_{q} $ as solutions from the Wannier equation \cite{Kochbuch,Kira2006,gunnar_prb,brem2019phonon,malic2018dark}, the electron (hole) operator $a^{c(v)\dagger}$ and relative $q$ and center of mass $Q$ momenta. They can be translated into electron and hole momenta via $q=\alpha k_e + \beta k_h$ and $Q=k_h-k_e$ with $\alpha(\beta)=\frac{m_e(m_h)}{m_e+m_h}$. By introducing the exciton operator in the pair-space, we can define an excitonic Hamilton operator including the interaction with phonons and a magnetic field. The Hamiltonian reads $H=H_{\text{0}}+H_{\text{x-phot}}+H_{\text{x-phon}}+H_{\text{x-magn}}$, where  $H_0=\sum_{Q,i}
    \eps^{i}_Q
    X^{\dagger i}_{Q} X^{i}_{Q}
    + \sum_{Q,\sigma}
    \hbar \omega^\sigma_Q
    c^{\dagger \sigma}_{Q} c^{\sigma}_{Q}
    +\sum_{Q,\zeta}
    \hbar \Omega^{\zeta}_Q
    B^{\dagger \zeta}_{Q} B^{\zeta}_{Q}$
is the interaction-free part for excitons, photons and
phonons with the excitonic energy $ \eps^i_Q$ in the state $i=(s_i,\eta_i)$ with the spin $s_i={\up\up,\up\down, \down\up, \down\down}$  and the valley $\eta_i=(\text{KK, KK', K}\Lambda, \Gamma \text{K})$, the photon energy $\hbar\omega_Q^{\sigma}$ with the polarization mode $\sigma$, and the phonon energy $\hbar\Omega_Q^\zeta$ with the phonon mode $\zeta$. The exciton-photon interaction reads $H_{\text{x-phot}}= \sum_{Q,i,\sigma} M^{i\sigma}_{Q}
    c^{\dagger \sigma}_{Q} X^{i}_{Q} + \text{h.c.}$ with the optical matrix element $ M^{i\sigma }_{Q}$ \cite{katsch2018theory,brem2018exciton}. 
Finally, the interaction between excitonic spin states $i,j$ and an in-plane magnetic field $B$ is described by the Hamiltonian
\begin{equation}
H_{\text{x-magn}}= \sum_{Q,i,j} 
   G^{i j} \frac{\mu_B}{2} B
    X_{Q}^{\dagger i} X_{Q}^{j}.
\end{equation}
The matrix element $G^{ij}$ reads in excitonic basis 
$G^{ij} = 
\left(g^c_{ij} \delta_{s_i^h,s_j^h}-g^v_{ji} \delta_{s_i^e,s_j^e} \right)
\sum_{q} \varphi_q^{i} \varphi_q^{j*} 
$ 
with the electrons (holes) keeping their spins, while mixing of spins  in the valence (conduction) band of one valley takes place, i.e. $g_{ij}^{c(v)} = g^{c(v)}_{\eta_i} \delta_{\eta_i,\eta_j} (1-\delta_{s_i^{e(h)},s_j^{e(h)}})$. Here, $ g^{c(v)}_{\eta_i}$ is the experimentally accessible g-factor for the conduction (valence) band in the valley $\eta_i$ and $\mu_B$ is the Bohr magneton. The g-factors in 2D materials are an ongoing topic of research \cite{forste2020exciton,wang2015magneto,mitioglu2015optical,koperski2018orbital} and still under debate as they can differ significantly for bright, dark and charged states. However, for the scope of this work, as we are interested in the qualitative behavior of dark states under magnetic fields, we  assume $ g^{c(v)}_{\eta_i}\approx 4$  for all states and materials (experimental values for the bright state are $g_{\text{MoSe}_2}=4.2, g_{\text{WSe}_2}=4.3, g _{\text{WS}_2}=4.0$ \cite{koperski2018orbital}).

Now, we have all ingredients at hand to derive the equation of motion for our key quantity, the photon-assisted polarization  $\braket{c_q^\dagger X_q^b}$ providing access to the PL, cf. \eqref{eq:pl_intensity}.
 However, the equation can be simplified resulting in an intuitive Elliott-like formula including both phonon- and magnetic field-induced PL. To get there, we perform a unitary transformation to include the magnetic field into $H_0$ and subsequently apply a cluster expansion approach to account for phonon-assisted radiative recombinations. 
We start by modifying the system with an unitary
transformation, such that $H_{\text{x-magn}}$ becomes included in $H_0$. To illustrate the idea, we simplify our system for now and assume an excitonic state $i=(s_i,\eta)$ with $Q\approx 0$ with different spins $s_i,s_j=\up\up,\up\down$ but same valley $\eta_i=\eta_j$. 
We can decouple the appearing spin-up and spin-down states and decompose the appearing matrix element $G^{ij}$ into a spin-subspace with Pauli matrices, and we find the field-induced mixing (neglecting for the moment the impact of  photons and phonons)
\begin{eqnarray}
  H =\sum_{s_i,s_j} 
   \eps^{\eta}_{s_i}
    X^{\dagger \eta}_{s_i} X^{\eta}_{s_i} + 
   G_{s_i s_j}^\eta \frac{\mu_B}{2} B
    X_{s_i}^{\dagger \eta} X_{ s_j}^{\eta}
=\nonumber
  \sum_{i}  \tilde \eps_{i}^{\eta}  \tilde X^{\eta\dagger}_{i}  \tilde X_{i}^\eta
\end{eqnarray}
with the new quantum number $i$, new energy 
$
 \tilde \eps_{i}^{\eta} = \eps_{i}^{\eta} 
  \pm \sqrt{\left( \frac{\Delta^{\eta}}{2} \right)^2 + \left( \frac{g}{2} \mu_B B \right)^2},
$
and the new state operator
$
  \tilde X_{i}^{\eta(\dagger)}
  = U_{i1}^{\eta} X_{\up\up}^{\eta (\dagger)} + U_{i2}^{\eta} X_{\up\down}^{\eta(\dagger)}
  = \frac{1}{\sqrt{1 + h_i^{\eta 2}}} \left( h_i^\eta X_{\up\up}^{\eta(\dagger)} + X_{\up\down}^{\eta(\dagger)} \right)
$, 
where we used the abbreviations $h_i^\eta = \Delta^{\eta} / (g \mu_B B) + \lambda_i \sqrt{1 + \Delta^{\eta 2} / (g \mu_B B)^2}$,  $\Delta^{\eta} = \eps_{\up\up}^\eta - \eps_{\up\down}^\eta$ and $\lambda_i=\pm$. The latter  can be positive or negative depending on the energetic ordering in the investigated TMD material. 
Note that for the definition of the transformation matrix $\hat U^{\eta}$, we exploited
$\tilde X^{\eta \dagger}_i = \hat U^{\eta} X^{\eta\dagger}_i= \sum_j U_{ij}^{\eta} X^{\eta \dagger}_j$ and $\hat U^{\eta \dagger} \hat U^{\eta} = \I$, i.e. a uniform transformation. 

We can now transform the rest of the Hamilton operator into this basis,  yielding for the exciton-photon coupling
$
  H_{\text{x-phot}} = \sum_i \tilde M_i  c \tilde X_i^{\eta\dagger}  +\text{h.c.}
$
with the new matrix elements $\tilde M_i = M U_{1i}$. Neglecting for the moment the interaction with phonons (hence $q \approx 0$), we derive the equation of motions for the photon-assisted polarization  appearing in \eqref{eq:pl_intensity}:
\begin{multline}
\label{photonEq}
 i \hbar \partial_t \braket{c^{\dagger} \tilde X^\eta_i} = (\tilde \eps^\eta_i - \hbar \omega) \braket{c^{\dagger} \tilde X^\eta_{i}}
  - \tilde M^{\eta}_i \braket{\tilde X^{\dagger \eta}_i \tilde X^{\eta}_i}.
\end{multline}
To calculate the PL intensity in presence of a magnetic field we solve this equation in the adiabatic limit \cite{brem2019phonon,Kochbuch, Kira2006} yielding
\begin{equation}
  \label{eq:plSpinForbidden1}
    I(\omega)
    \propto \hbar \omega \Im
\sum_{i\eta}
      \frac{ \big(\tilde M^{\eta}_i \big)^2 N_{i}^{\eta}} {\tilde \eps^\eta_i  - \hbar \omega - i \gamma^{\eta}_i} 
\end{equation}
with the exciton occupations $N_{i}^{\eta}$, excitonic energy $\tilde \eps^\eta_i$ and dephasing $\gamma^{\eta}_i$, where $\eta$ and $i$ are the exciton and spin index. \\

{\noindent\textbf{Brightening of spin-dark excitons:}} For a better understanding of the influence of the magnetic field, we disregard for the moment the impact of momentum-dark excitons and consider only  the bright state $\eta=\text{KK}, i={\up\up}$ (denoted by B)  and the spin-dark state $\eta=\text{KK}, i={\up\down}$ (denoted by D). Furthermore we consider the situation $\mu_BB\ll\Delta^\eta$, where the energy difference between spin-allowed and spin-dark state is large compared to the Zeeman splitting. Here, the Zeeman term gives only a small correction to the energy, which is the case in tungsten-based TMDs for experimentally available magnetic field strengths. Then, we split up the sum over $\eta$ in  \eqref{eq:plSpinForbidden1} into dark and bright state and enter the solutions $\tilde M^{\eta}_i = M U^{\eta}_{1i}$. Moreover, we set $U^{\eta} =\text{const}$ for the bright states and perform a Taylor  expansion of $U^{\eta}$ for the dark states. This results in a more intuitive expression for the PL 
\begin{equation}
  \label{eq:plSpinForbidden}
  \begin{split}
    I
    \propto \hbar \omega \Im
    &\left[
      \frac{M^2 N_{B}} {\eps_B - \hbar \omega - i \gamma_B} 
      +  
       \frac{M^2\left(\frac{g \mu_B B}{2 \Delta^{\text{KK}}} \right)^2N_D} {\eps_{D} - \hbar \omega - i \gamma_D}
    \right].
  \end{split}
\end{equation}
The first term  describes the direct PL contribution stemming from the bright KK$^{\up\up}$ exciton and resulting in a resonance  at the energy $\eps_B$. In addition, the magnetic field appearing in the second term gives rise to a new resonance at the position $\eps_D$ due to the activation of spin-dark excitons. Assuming that excitons quickly  thermalize,  exciton densities can be approximated by equilibrium Boltzmann distributions \cite{selig2018dark,brem2019phonon}.

\begin{figure}[t]
\includegraphics[width=\linewidth]{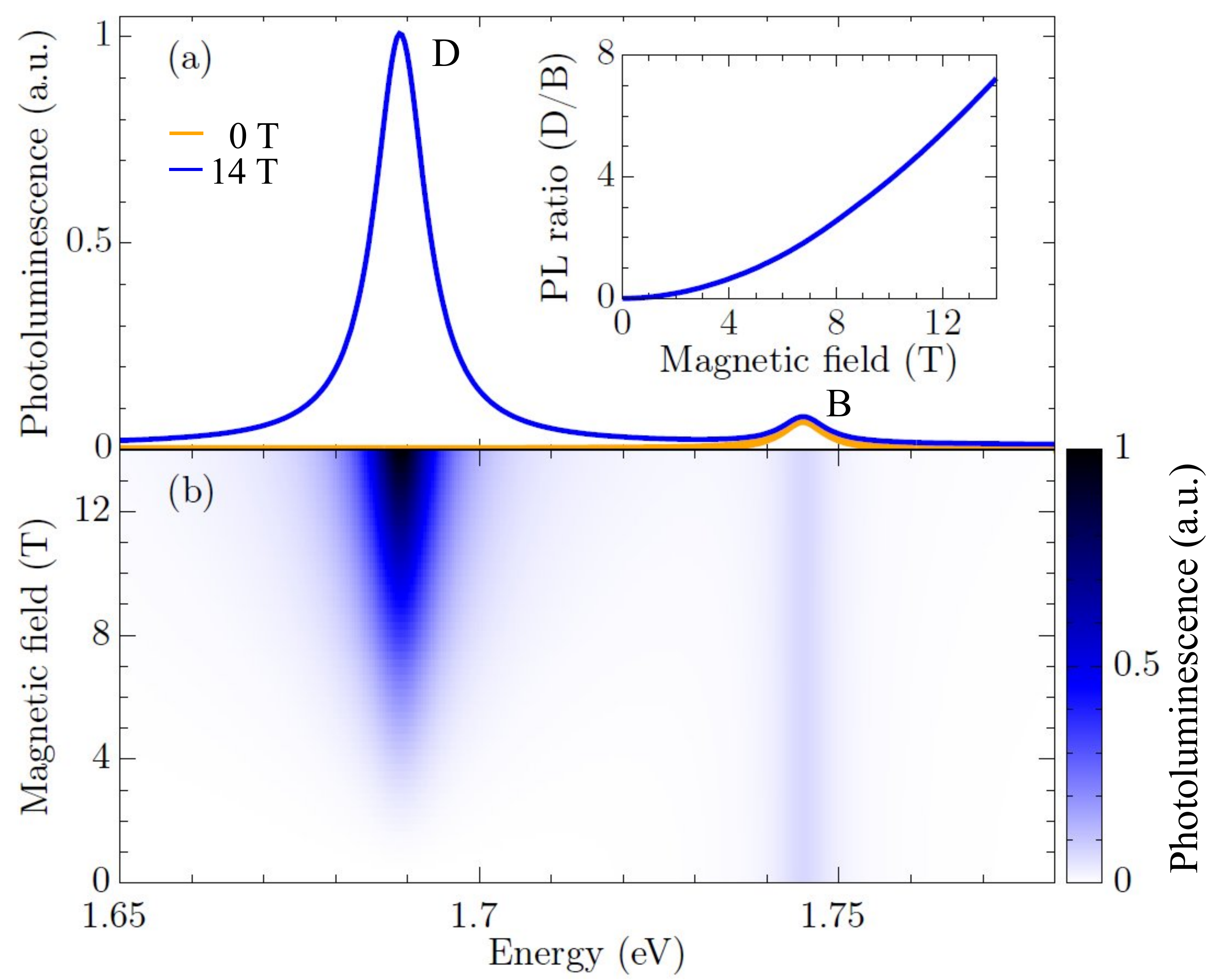} 
    \caption{\textbf{Brightening of spin-dark excitons.} Photoluminescence of hBN-encapsulated WSe$_2$ monolayer at an exemplary low temperature of 35 K with and without a magnetic field. Note that exciton-phonon scattering and phonon-sidebands have been neglected in this study.  (a) While the PL at 0 T  is determined by the bright $\text{KK}^{\up\up}$ exciton at 1.745 eV,  at 14 T the PL is clearly dominated by a new resonance at 1.69 eV stemming from the spin-dark $\text{KK}^{\up\down}$ exciton. The inset reveals a quadratic behavior for the dark-bright peak ratio. (b) Surface plot of the PL as a function of energy and magnetic field strength. Note that the spectra are normalized  to the bright exciton peak. The spin-dark exciton becomes visible from approximately 2 T on and grows with the magnetic field. 
 }
   \label{figure2}
\end{figure}

 Now, we numerically evaluate \eqref{eq:plSpinForbidden} to calculate the PL spectrum of  WSe$_2$ as an exemplary TMD material. 
Figure \ref{figure2}(a) shows the PL spectrum
 at a temperature of 35 K with (blue) and without (orange) the magnetic field. 
The peak at 1.745 eV corresponds to the bright exciton, while the extremely pronounced peak at 1.69 eV stems from the brightened spin-dark exciton. 
The intensity of the peak is given by both the exciton occupation $N_D$ and the magnetic field strength assuming the same dephasing rates $\gamma_D=\gamma_B$. We find a quadratic B-field dependence for the PL intensity ratio between the dark (D) and the bright state (B) - in good agreement with experimental observations \cite{zhang2017magnetic,molas2017brightening}. This reflects the prefactor $(\frac{g \mu_B B}{2 \Delta} )^2$  in \eqref{eq:plSpinForbidden}. Here, TMD specific parameters, such as the energetic difference between the dark and bright state $\Delta$ and the g-factor, play an important role. 

Figure \ref{figure2}(b) shows a surface plot illustrating the dependence of PL on the magnetic field. The stronger the field, the more efficient is the brightening of the spin-dark exciton and the more pronounced is its PL peak.  Furthermore, for an in-plane magnetic field, the Zeeman shift is known to be negligible and therefore we do not observe a shift in energy \cite{molas2017brightening}. 
 The dark exciton peak starts  to appear for magnetic fields B$>$2 T at the considered exemplary temperature of 35 K. Exploiting \eqref{eq:plSpinForbidden}, we find an analytic expression for the intensity ratio $
  \text{I(D)/I(B)}
  =    \xi(\text{TMD,T}) \,  B^2
$
which reveals the quadratic behavior shown in the inset of Fig. \ref{figure2}(a). 
The curvature of the parabola $\xi(\text{TMD,T}) =\frac{\mu_B}{2} \frac{\eps_D}{\eps_B}
  \frac{N_D(T)} {N_B(T)}  \frac{g}{\Delta} $ is TMD-specific and depends specifically on the (i) relative energetic position of dark and bright excitons ($\frac{\eps_D}{\eps_B}\approx 1$), (ii) the TMD-specific g-factor and dark-bright energy splitting $\Delta$, and (iii) the relative temperature-dependent exciton densities ($\frac{N_D} {N_B}  = e^{-\frac{(\eps_D - \eps_B)}{k_BT}}$). In materials with $\eps_D > \eps_B$ (e.g. MoSe$_2$) this factor will quickly approach 0 at low
temperatures and hence we do not expect the dark state to brighten up, since its occupation is very low. On the other side,  for tungsten-based materials  with  $\eps_D < \eps_B$   the factor will instead approach infinity at low temperatures and enables pronounced brightening of dark states.
However, as the temperature increases the factor $\frac{N_D} {N_B}$ becomes smaller and hence the intensity of the peaks decreases as dark excitons thermalize and recombine via phonons. This is why spin-dark states are observed mainly at low temperatures \cite{molas2017brightening,zhang2017magnetic}.
For the curvature we can extract $\xi(\text{WS}_2\text{,35 K})=0.0018 \, \text{T}^{-2}$ and  $\xi(\text{WSe}_2\text{,35 K})=0.0038 \, \text{T}^{-2}$. This means that WSe$_2$ is by a factor of two more responsive to an external magnetic field,  which is in good agreement with experimentally observed values \cite{molas2017brightening}. Note that slight differences in the g-factor of the two materials are expected to slightly change the difference in the curvature.\\

{\noindent\textbf{Brightening of spin- and momentum-dark excitons:}}
So far, we have only included the effect of phonons in the linewidth of the exciton resonances \cite{selig2016excitonic,brem2018exciton,brem2019phonon}. However, under certain circumstances phonons can drive indirect radiative transitions from  momentum-dark states  \cite{brem2019phonon,lindlau2018role}.  
Hence, to fully understand the influence of the magnetic field on PL spectra, we now  include exciton-phonon scattering and consider both momentum- and spin-dark exciton states. 

We extend the exciton Hamiltonian by the exciton-phonon coupling 
\begin{equation}
H_{\text{x-phon}}= \sum_{\substack{Q,Q'\\ \eta_1, \eta_2, l,j,\zeta }}
    D^{\eta_1 l\eta_2j\zeta}_{Q'}
    \tilde{X}^{\dagger \eta_1,l}_{Q + Q'} \tilde{X}^{\eta_2,j}_{Q} B_{-Q'}^{\dagger \zeta}
    + \text{h.c.}
\end{equation} 
which creates an exciton with the momentum $Q+Q'$ in the state $\eta_1,l$ and annihilates an exciton with the momentum $Q$ in the state $\eta_2,j$  under creation of a phonon with the momentum $-Q'$ and the mode $\zeta=(\text{LA,TA,LO,TO})$.
The exciton-phonon matrix element in the magnetic field basis reads $D^{\eta_1l \eta_2j\zeta}_{ Q'} = 
\sum_{k,\lambda,n,i}
 U^{\eta_1}_{nl} \left( \varphi_{k}^{\eta_1*} d^{\lambda}_{Q'\zeta} \varphi_{k+ \xi^\lambda Q'}^{\eta_2}  \right) U^{\eta_2 *}_{ij}
$ with transformation matrices $U^{\eta_1(\eta_2) *}_{nl(ij)}$. Here,  $d^{\lambda}_{Q\zeta}$ denotes the electron-phonon coupling elements in the band $\lambda = (c,v)$ \cite{brem2019phonon}. Furthermore, we introduced $\xi^\lambda=\alpha,\beta$ with excitonic mass factors $\alpha(\beta)=\frac{m_e(m_h)}{m_e+m_h}$ for the valence(conduction) band.
We can see that the phonon introduces a momentum $Q'$ into the system, which is transferred to the exciton. Depending on the efficiency of the exciton-phonon coupling, this can lead to the activation of momentum-dark excitons \cite{brem2019phonon}. 

The transformation into the magnetic field basis enables us to exploit the TMD Bloch equations for phonon-assisted photoluminescence derived in our previous work \cite{brem2019phonon} with the modified optical matrix element $\tilde M^{\eta}_i \rightarrow \tilde M^{\mu} $ and exciton energies $\tilde \eps^\eta_i + \frac{\hbar^2 Q^2}{2m^{\eta i}} \rightarrow \tilde \eps^{\mu}_Q  $ with the compound index $\mu=\eta,i$. We obtain a new expression for direct photoluminescence
\begin{equation} \label{PLB} 
 I(\omega)^{b}\propto
\sum_{\mu} \dfrac{|\tilde M^{\mu} |^2  \gamma^{\mu}_{\mathbf{0}} N^{\mu}_\mathbf{0}}{(\tilde \eps^\mu_\mathbf{0}-\omega)^2+(\gamma^{\mu}_{\mathbf{0}}+\Gamma^{\mu}_\mathbf{0})^2} ,
\end{equation}
which is analogue to \eqref{eq:plSpinForbidden1}   but now not only includes the radiative dephasing $\gamma^{\mu}_{\mathbf{0}}$ but also phonon-induced dephasing $\Gamma_\mathbf{Q}^{\mu}$. For the indirect phonon-assisted  photoluminescence allowing us to reach momentum-dark excitons we obtain the expression
\begin{equation}\label{PLD}
 I(\omega)^{d} \propto 
  \sum_{\mu\nu\mathbf{Q},\zeta\pm} 
\Omega^\mu(\omega)
 \dfrac{|\tilde{D}^{\mu\nu}_{\zeta\mathbf{Q}}|^2 N^\nu_\mathbf{Q} \eta^{\pm}_{\zeta\mathbf{Q}} \Gamma^\nu_\mathbf{Q}}{( \tilde \eps^{\nu}_\mathbf{Q}\pm\Omega^\zeta_\mathbf{Q} -\omega)^2 +(\Gamma^\nu_\mathbf{Q})^2},
\end{equation}
where we have introduced the abbreviation $ \Omega^\mu(\omega)= \frac{|\tilde M^{\mu} |^2}{(\tilde \eps^\mu_\mathbf{0}-\omega)^2+(\gamma^{\mu}_{\mathbf{0}}+\Gamma^{\mu}_\mathbf{0})^2}   $ determining i.a. the oscillator strength. 
The position of the new phonon-induced signatures in the PL is determined by the energy of the exciton $\tilde \eps^{\nu}_\mathbf{Q}$ and  the energy of the involved phonon $\pm\Omega^\zeta_\mathbf{Q}$. The sign describes either the absorption (+) or emission (-) of phonons. We take into account all in-plane optical and acoustic phonon-modes. Moreover, the appearing phonon occupation $\eta_{\zeta Q}^{\pm} = \left(
\frac{1}{2} \mp \frac{1}{2} + n^{\text{phon}}_{\zeta Q}
\right)$ is assumed to correspond to the Bose equilibrium distribution according to the bath approximation \cite{axt1996influence}. Since dark states can not decay radiatively, the peak width is only given by non-radiative dephasing processes $\Gamma^{\nu}_{\bf Q}$. 
The total photoluminescence in presence of phonons and magnetic fields is obtained by adding \eqref{PLB} and \eqref{PLD}, which now includes mixing of spin and momenta by the appearing sums $\mu=\text{KK}^{\up\up},\text{KK}^{\up\down}$ and $\nu=\text{KK}^{\up\updownarrow},\text{KK'}^{\up\updownarrow},\text{K}\Lambda^{\up\updownarrow},\Gamma \text{K}^{\updownarrow \up}, \Gamma \text{K}^{\updownarrow \down}$.

\begin{figure}[t]
  \begin{center}
\includegraphics[width=\linewidth]{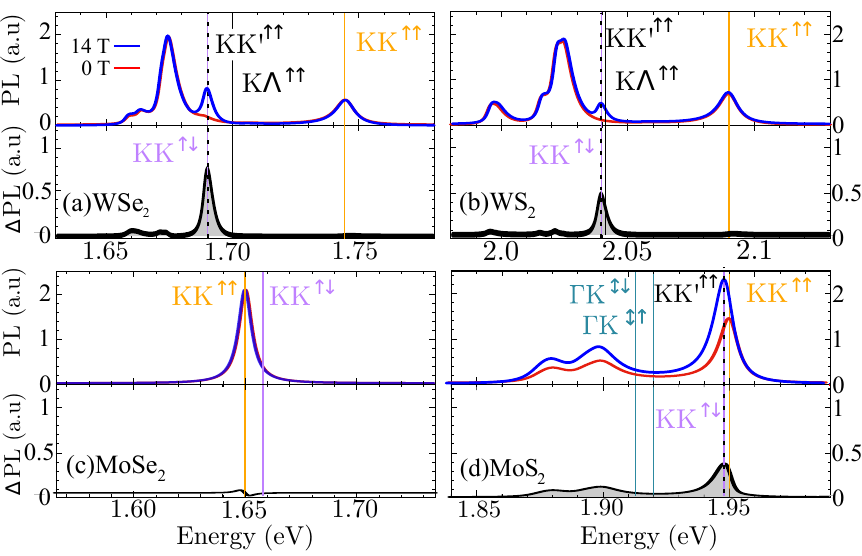} 
\end{center}
    \caption{\textbf{Brightening of spin- and momentum-dark states.}
PL with and without a magnetic field studied at 35 K for different hBN-encapsulated TMD monolayers. The lower panels show the PL difference $\Delta \text{PL}$ = $I_{14 \text{ T}} - I_{0 \text{ T}}$ illustrating directly the impact of the magnetic field. Without the latter, the PL in all TMDs but MoSe$_2$ is dominated by phonon-sidebands  of momentum-dark  $\text{KK'}^{\up\up}$,  $K\Lambda^{\up\up}$ or $\Gamma \text{K}^{\up\up}$ excitons. They are located approx. 50-100 meV below the bright $\text{KK}^{\up\up}$ state.  In presence of a magnetic field, additional shoulders are observed at the position of  spin-dark $\text{KK}^{\up\down}$ or below the $\Gamma \text{K}^{\up\down,\down\up }$ excitons. The latter state appearing in MoS$_2$ is both spin- and momentum-dark and requires brightening via both phonons and an in-plane magnetic field. }
   \label{figure3}
\end{figure}

Now, we investigate the PL spectra at an exemplary low  temperature of 35 K  for both tungsten- and molybdenum-based TMDs, cf. Fig. \ref{figure3}. We directly compare the spectrum with  and without the presence of a magnetic field. The lower panel of each picture shows the differential PL directly illustrating the impact of the field.
The first observation is that even without the magnetic field additional low-energy signatures are observed in most TMD materials. They stem from momentum-dark states (black vertical lines) and are activated via phonon-assisted radiative recombination \cite{brem2019phonon}. 
In the case of  WSe$_2$, the peaks between 1.65 and 1.70 eV stem from phonon emission and absorption from energetically lower lying KK'$^{\up\up}$ and K$\Lambda^{\up\up}$ excitons, respectively, cf. the red line in Fig. \ref{figure3}(a). 
 Those phonon-sidebands appear since phonons add an additional center-of-mass momentum allowing excitons to recombine. The features are not observed directly at the position of these excitons, but are shifted by the energy of the involved phonon, i.e. 
 $\tilde \eps^{\nu}_\mathbf{Q}\pm\Omega^\zeta_\mathbf{Q}  $ (see \eqref{PLD}), where $+$ is phonon absorption and $-$ phonon emission.
      At low temperatures, phonon emission is dominant and hence the phonon sidebands are located 15-50 meV below the exciton position, corresponding to the phonon energies of LO/LA and TO/TA phonons. For WS$_2$, we obtain a similar picture. The differences can be explained by different phonon energies and occupations and energetic positions of momentum-dark KK'$^{\up\up}$ and K$\Lambda^{\up\up}$ excitons (cf. black vertical  lines). Note that in principal all excitonic states exhibit phonon sidebands but only the energetically lowest one have a sufficient occupation to be visible in PL. 
 
As MoSe$_2$ does not exhibit any energetically lower lying dark excitonic states \cite{andor,malic2018dark}, its PL  is dominated by the bright KK$^{\up\up}$ exciton, cf. Fig. \ref{figure3}(c).
For MoS$_2$, the energetically lowest state is the momentum-dark $\Gamma$K which is degenerated and exhibits energetically close lying spin-allowed and spin-forbidden $\up\up,\up\down,\down\up \text{ and } \down\down$ states.
This degeneracy leads, in combination with overlapping phonon replica, to broader low energy peaks between 1.86 and 1.91 eV, cf. Fig. \ref{figure3}(d).  Moreover, the lower energy shoulder of the bright peak can be explained by phonon replica of the  $\text{KK'}^{\up\up}$ exciton. Note that the linewidths of the peaks are calculated on a microscopic level (for more details see our previous work \cite{selig2016excitonic,brem2018exciton,khatibi2018impact}) and are in good agreement with experiments \cite{molas2017brightening}. Since the phonon replica and linewidths are very sensitive to the exact position and contributions of the valleys \cite{khatibi2018impact}, the appearance of phonon sidebands can be a signatures of lower lying $\Gamma$K excitons in MoS$_2$.

Now, we investigate the changes of PL signatures in presence of a magnetic field, cf. blue lines in Fig. \ref{figure3} and for a better illustration the differential PL spectra $\Delta \text{PL}$ = $I_{14 \text{ T}} - I_{0 \text{ T}}$ in the lower panels of the figure.
We observe for both  tungsten-based TMDs and MoS$_2$ an upcoming peak around 50-70 meV below the bright KK$^{\up\up}$. We can trace back this new peaks to the activation of  (i) spin-dark KK$^{\up\down}$ excitons in WSe$_2$ and WS$_2$ , and (ii) spin- and momentum-dark $\Gamma$K$^{\uparrow\downarrow}$ and $\Gamma$K$^{\downarrow\uparrow}$ states in MoS$_2$. Moreover, MoS$_2$ exhibits an additional peak just below the bright KK exciton which can be assigned to the spin-dark KK$^{\up\down}$ exciton. In contrast, MoSe$_2$ does not exhibit any additional field- or phonon-induced peaks.  This reflects the excitonic landscape in this material with the bright KK$^{\up\up}$ as the energetically lowest state \cite{malic2018dark,deilmann2019finite}. Note that very recent experiments by Lu et al. \cite{lu2019magnetic} and Robert et al. \cite{robert2020measurement} observed an upcoming peak 1-2 meV above the bright peak in MoSe$_2$  in very high magnetic field (B$\approx$30-60 T), suggesting a possible brightening of KK$^{\up\down}$ excitons in these materials. 
Note that we find small shifts at the position of the KK$^{\up\up}$ exciton reflecting the Zeeman shift. It is in the range of $10^{-1}$ meV and only visible in the differential PL spectra.

The field-induced difference in the PL  between tungsten- and molybdenum-based TMDs stems from different underlying brightening mechanisms: While in WSe$_2$ and WS$_2$ the magnetic field induces a mixing of spin-allowed and spin-forbidden KK excitons, resulting in a peak at the position of the spin forbidden KK$^{\up\down}$ exciton, in MoS$_2$ it additionally couples spin-allowed and momentum-dark $\Gamma\text{K}^{\up\up}$ states with spin-\text{and} momentum-dark $\Gamma\text{K}^{\down\up,\up\down }$ excitons, resulting in phonon replica energetically below these $\Gamma\text{K}^{\down\up,\up\down }$ states.
The strong mixing of spin- and momentum-dark states in MoS$_2$ can be traced back to the degeneracy of the states and the strong electron-phonon matrix elements \cite{jin2014intrinsic,kaasbjerg2012phonon}. Since there is no splitting of spin-states in the valence band of the $\Gamma$ valley and  since the splitting is small in the conduction band (in the range of 3 meV) \cite{andor}, $\Gamma$K excitons are energetically very close enhancing the interaction and mixing of these states. Note that the splitting in the $\Gamma$ states is more pronounced in the excitonic picture, i.e. $\eps_{\Gamma\text{K}}^{\up\up}-\eps_{\Gamma \text{K}}^{\up\down}= 8\text{ meV}$ due to the influence of different exciton masses.
  In tungsten-based TMDs, the spin- and momentum-dark KK'$^{\up\down}$ and K$\Lambda^{\up\down}$ states are energetically higher than the bright state \cite{andor,malic2018dark,deilmann2019finite} and hence do not contribute to the PL due to the very low occupation.

Comparing our results with experimental observations \cite{molas2017brightening}, we find a very good qualitative agreement  of the field-dependent PL in all four TMDs. Both theory and experiment find an additional narrow peak in tungsten-based TMDs, a broad low-energy peak in MoS$_2$ and no field-induced signatures in MoSe$_2$ in the investigated magnetic fields of up to 15 T. Note that in the experiment a peak splitting of the spin-dark resonance appears in the presence of a magnetic field, which can be ascribed to the Coulomb exchange interaction \cite{molas2017brightening} that has not been taken into account in our model. 
Since the exchange interaction only affects momentum-allowed states \cite{selig2019quenching}, the splitting does not occur for MoS$_2$, where spin \textit{and} momentum-dark $\Gamma$K$^{\up\down}$ excitons play the crucial role.

\begin{figure}[t!]
  \begin{center}
\includegraphics[width=\linewidth]{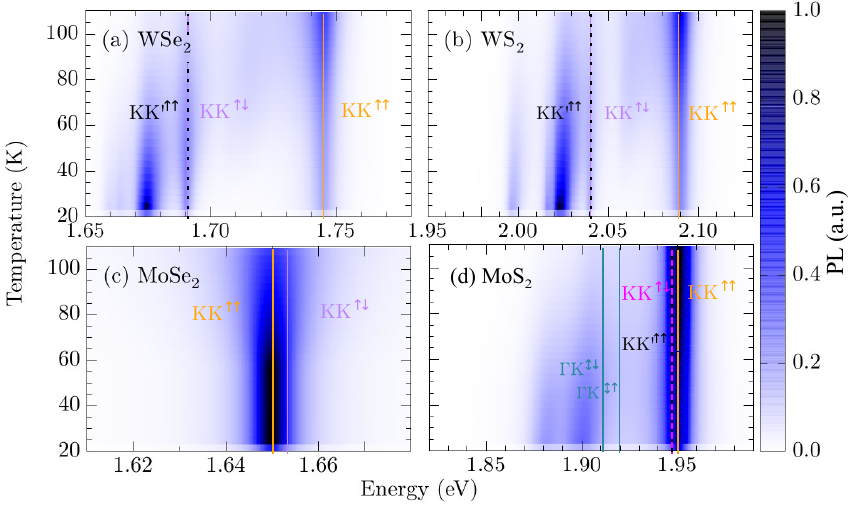} 
\end{center}
    \caption{\textbf{Temperature dependence of PL in magnetic field.} 
Temperature dependence of photoluminescence in different TMD monolayers at the fixed magnetic field of 14 T. Dark exciton signatures can be observed  up to 100 K in  tungsten-based materials. In MoS$_2$, the resonances are much broader due to the overlap of phonon-sidebands of spin-and momentum-dark $\Gamma$K excitons. }
   \label{figure4}
\end{figure}

So far we have discussed PL signatures of dark states at one exemplary temperature. Now, we vary the temperature and investigate how the impact of the magnetic field and phonons changes, cf. Fig. \ref{figure4}. We assume a constant magnetic field of 14 T for all investigated TMDs. We find for  tungsten-based TMDs that  at low temperatures, the lowest resonances stemming from  KK$'^{\up\up}$ and KK$^{\up\down}$ excitons dominate the PL spectrum, while at temperature above 60 K the bright peak becomes crucial - in agreement with  experimental results \cite{robert2017fine}. Note that the intensity dependence on the temperature is a result of an interplay between phonon and exciton occupations in the corresponding exciton state on the one side and the exciton-phonon scattering  determining the linewidth of the resonance on the other side \cite{selig2016excitonic,brem2018exciton,brem2019phonon}.

For  molybdenum-based materials, we find that MoSe$_2$ is dominated by the bright KK$^{\up\up}$ exciton at all temperatures, however exhibiting an increased peak broadening at higher temperatures due to the enhanced exciton-phonon interaction. 
In contrast, MoS$_2$ shows even at low temperatures two broad peaks stemming from (i) phonon sidebands of the spin- and momentum-dark $\Gamma$K$^{\up\down,\down\up}$ excitons and (ii) direct emission from the momentum-allowed KK$^{\up\downarrow}$ exciton around 1.95 eV. With increasing temperature, the indirect peaks become broader and decreases in intensity due to enhanced exciton-phonon scattering.\\

{\noindent \textbf{Conclusions:}} We have investigated the impact of an in-plane magnetic field on  optical properties of transition metal dichalcogenides. Exploiting a fully quantum-mechanical and microscopic approach, we provide insights into signatures of  momentum- and spin-dark excitons in photoluminescence spectra. We find that the field-induced mixing of spin-up and spin-down states results in a brightening of  spin-dark excitons resulting in new resonances.  We show that the origin of the new peak is the direct emission of the KK$^{\up\down}$ exciton in tungsten-based materials, whereas for MoS$_2$ it is the indirect, phonon-induced transition from the spin- and momentum-dark $\Gamma$K$^{\downarrow\uparrow,\up\down}$ excitons.  Our work provides microscopic insights into experimentally observed photoluminescence spectra in presence of magnetic fields and overall sheds light on the excitonic landscape in 2D materials. \\

{\noindent \textbf{Acknowledgments:}} This project has received funding from the Swedish
Research Council (VR, project number 2018-00734) and the European Union's Horizon
2020 research and innovation program under grant
agreement No 881603 (Graphene Flagship). Furthermore, we acknowledge support by  the Chalmers Area of Advance in Nanoscience and Nanotechnology.

\bibliographystyle{apsrev4-1}
%
\end{document}